\documentclass[10pts,twocolumn]{article}
\usepackage[dvips]{graphics}
\usepackage{epsfig}
\usepackage[centertags]{amsmath}
\usepackage{amsfonts}
\usepackage{amssymb}
\usepackage{amsthm}
\usepackage{newlfont}

\begin{document}
\title{ A sphere moving down the surface of a static sphere and a simple phase diagram.}
\author{V. Jayanth, Mechanical Engineering; C. Raghunandan, EEE; \\
Anindya Kumar Biswas, Physics;\\
BITS-Pilani Goa Campus, Vasco, Goa-403726.\\
email:anindya@bits-goa.ac.in}
\date{\today}
\maketitle
\begin{abstract}
A small sphere placed on the top of a big static frictionless
sphere, slips until it leaves the surface at an angle
$\theta_{l}=\cos^{-1}{2/3}$. On the other extreme, if the surface
of the big sphere has coefficient of static friction, $\mu_s
\rightarrow\infty$, the small sphere starts rolling and continues
to do so until it leaves the surface at an angle $\theta_{l}
=\cos^{-1}{10/17}$. In the case where, $0<\mu_s<\infty$, we get a
simple phase diagram. The three phases are pure rolling, rolling
with slipping and detached state. One phase line separates pure
rolling from rolling with slipping. This diagram is obtained when
stopping angles for pure rolling are plotted against static
friction coefficients $\mu_s$. Study in this article is restricted
to the case when the mobile sphere starts at the top of the static
sphere with infinitesimal kinetic energy.
\end{abstract}
\begin{section}{Introduction}
An interesting text book problem in the introductory
mechanics(\cite{res}, \cite{klep}, \cite{hite}) is to determine
the release point for a block of mass m. The block slides without
friction on the surface of a static sphere of radius R. The block
starts from "rest", i.e. with infinitesimal velocity, from the top
of the sphere. The block looses contact with the
 sphere, by the time it travels a vertical distance
$\frac{R}{3}$. This is equivalent to an angular displacement
$cos^{-1}\frac{2}{3}$. This result draws attention. The angular
displacement does not depend either on the mass of the block or,
the radius of the sphere. The result is general.\\
\indent When there is friction on the surface of a sphere
(\cite{Mele}, \cite{Munga}, \cite{Lange}), the block leaving with
infinitesimal velocity at the top of the static sphere,
immediately stops. This dull situation changes to better as soon
as the block is given a finite velocity at the top of the static
sphere. The block may now continue to slip. It may also get
released from the surface of the sphere below. Whether the block
will leave the sphere below or, stick depends on the amount of
friction. In that sense, the initial velocity and the static
friction start playing with the block in fixing the fate of it. In
an interesting report (\cite{Lange}), the authors got an exact
relation between initial velocity and the coefficient of static
friction, requiring the block to be released from the surface with
zero velocity. Once drawn by them(\cite{Lange}), two features
appeared. The plot resembles $tanh$ curve. The curve separates two
regions on the plot clearly. The lower region refers to the set of
possibilities when the block will stick. The upper region tells us
when the
block will leave.\\
\indent Once the block is replaced by a sphere, two things occur.
In the case the coefficient of static friction is
zero(\cite{Spiegel}), the mobile sphere slides and leaves the
surface exactly as the block does. When the coefficient of static
friction is very large(\cite{Spiegel}) and tends to infinity, the
mobile sphere instead of stopping immediately as in the case of
the block, starts doing something else. It starts to pure roll. It
pure rolls until it leaves the surface. As the coefficient of
static friction is retained intermediate between these two extreme
limits, zero and infinity, feasibility of a new kind of motion
arises. The sphere pure rolls initially, then starts skidding
along with rolling (\cite{Bach}, \cite{Song}). This going over to
mixed rolling happens at a particular angle. The mobile sphere,
once trapped in the mixed rolling phase, continues to do so until
it
leaves the surface of the static sphere.\\
\indent Incidentally, the cross-over angle bears a simple
relationship with the coefficient of static friction. This refers
to cross-over from pure to mixed rolling. Once the cross-over
angle is plotted, against the coefficient of static friction, two
features appear. The plot bears resemblance with $tanh$ curve. The
curve separates two regions on the plot clearly. The lower region
refers to the set of possibilities when the mobile object sticks
to the pure rolling phase. The upper region refers to the mixed
rolling phase. These qualitative features of the phase diagram
seem to be general, independent of the nature of the movable
object, whether
it is a block, or, a sphere. \\
\indent The phase diagram in the case of Mele et.al.
(\cite{Lange}), is external. Both the axes, initial velocity and
coefficient of static friction, refer to parameters controlled
from outside. In the case of a mobile sphere, the phase plot is
apparently partly internal. The parameter $\theta$ though may
appear internal, refers to the scale for potential energy, which
is tuned from below, just as the coefficient of static friction
is. If $v$ in the case of Mele et.al. (\cite{Lange}), determines
the initial kinetic energy given to the block, varying $\theta$ in
the case of riding sphere, implies going over from one potential
energy to another. Though we have only one initial kinetic energy
for a set-up, we have an infinite sequence of potential energies
for the same set-up, the sequence marked by $\theta$.\linebreak
\noindent As the initial velocity of the movable sphere, at the
top, is changed from zero to non-zero, a mixed rolling phase is
likely to appear in the resulting three dimensional, ($v$,
$\theta$,
$\mu_s$), phase diagram.\\
\indent In the section II, the motion of the small sphere at the
top of a static sphere is dealt with in the context of the
frictionless case. In the following section III, the angle at
which the mobile sphere leaves the static sphere, in the case of
infinite friction, is derived. In the section IV, the intermediate
coefficient of static friction is considered, with the depiction
of consequent phase diagram. In the subsection IV.1, the general
expression for the release angle is arrived at, followed by a
diagram of $\theta_{l}$ against $\theta_{0}$. In the next
subsection, we determine the order parameter and a critical
exponent associated with the phase transition at $\theta_{0}$.
\end{section}
\begin{section}{$\mu_s=0$}
\begin{figure}
\centering
\includegraphics{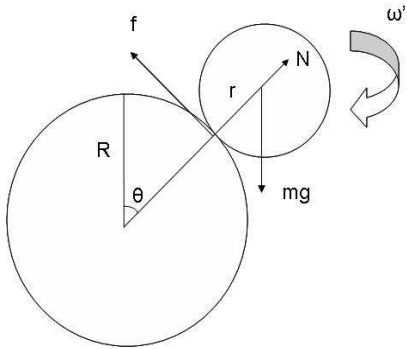}
\caption{\small{The small sphere is rolling down the big static sphere}}\label{Figure1}
\end{figure}
As the Fig.1 illustrates, three forces are acting on the small
sphere. The normal force, $N$, is acting through the contact
surface, along the line joining the centers. The weight of the
small sphere, $mg$, is passing through the center of mass
vertically downwards. The static frictional force is being exerted
by the static sphere below, through the surface of contact,
opposite to the direction of motion. The motion of the center of
mass of the small sphere, until it leaves the surface of the big
sphere is a circle about the center of the big sphere. This makes,
use of planar polar coordinate system for the description of
motion particularly handy, with the the center of the big sphere
as the origin and the top most position chosen as $\theta=0$.
Velocity of the center of mass, $v$, then reduces to
$(R+r)\omega$. Acceleration of the center of mass of the small
sphere, splits up into radial acceleration $\frac{v^{2}}{r+R}$ and
cross-radial acceleration $(R+r)\frac{d\omega}{dt}$. Newton's
second law yields radial and cross-radial equations of motion. The
radial equation of motion is,
\begin{equation}\label{e:ra}
mgcos(\theta)-N(\theta)=\frac{mv^{2}}{r+R}
\end{equation}
In this case, as friction is absent, integration of cross-radial equation of motion
with respect to $\theta$, leads to the
energy conservation equation,
\begin{equation}\label{e:en}
\frac{1}{2}mv^{2}-mg(r+R)(1-cos\theta)=0
\end{equation}
As the small sphere leaves the surface of the big sphere,
\begin{equation}\label{e:normal}
N(\theta)=0.
\end{equation}
This fact together with the equations (\ref{e:ra}) and (\ref{e:en}),
yields a quantitative expression for the release angle,
\begin{equation}
\theta=\cos^{-1}{2/3}=0.8411.
\end{equation}
\end{section}
\begin{section}{$\mu_s\rightarrow\infty$}
The small sphere, given a small perturbation, starts rolling
without slipping. The radial equation remains as before for the
$\mu_s=0$ case.
The energy conservation equation(\ref{e:en}) modifies to
\begin{equation}\label{e:enr}
\frac{1}{2}mv^{2}+\frac{1}{2}I(\omega^{\prime})^{2}-mg(r+R)(1-cos\theta)=0.
\end{equation}
where, $\omega^{\prime}$ refers to angular velocity with respect
to the center of mass of the small sphere. $I$ in the above
equation(\ref{e:enr}) corresponds to the moment of inertia for the
small sphere,
\begin{equation}
I = \frac{2}{5}mr^{2}.
\end{equation}
During pure rolling,
\begin{equation}\label{e:pr}
v = \omega^{\prime} r.
\end{equation}
This fact combined with the radial and energy conservation
equations, (\ref{e:ra},\ref{e:enr}) added with the equation (\ref{e:normal})
yields the angle of release
\begin{equation}
\theta =\cos^{-1}{10/17}= 0.9419.
\end{equation}
$\mu_s\rightarrow\infty$ guaranties that the small sphere
continues to pure roll without going over to slipping, until it
takes off from the surface of the sphere below.
\end{section}
\begin{section}{$0<\mu_s<\infty$}
The cross radial equation of motion in this case appears as,
\begin{equation}\label{e:cra}
ma_{\theta}= F_{\theta},
\end{equation}
with,
\begin{equation}
F_{\theta} = mg sin\theta - f,
\end{equation}
being the cross radial force.
The torque equation for rotation of the small sphere about its
center of mass, is
\begin{equation}\label{e:tor}
I\alpha^{\prime} = rf
\end{equation}
where, $f$, is the static frictional force acting on the small
sphere. This is acting tangentially at the point of contact with
the sphere below. Since, the velocity of the center of mass of the
small sphere is in the cross-radial direction, the equation
(\ref{e:pr}) for pure rolling can be written as,
\begin{equation}\label{e:purer}
a_\theta = r\alpha^{\prime}
\end{equation}
These three equations (\ref{e:cra}), (\ref{e:tor}),
(\ref{e:purer}), once combined, lead to a simple dependence of
angular position, $\theta$, on the frictional force, for the pure
rolling to persist,
\begin{equation}\label{e:f}
f= \frac{2}{7}mgsin\theta.
\end{equation}
Related matters are dealt with in a slightly different context in
the references (\cite{Bach}, \cite{Song}). As $\theta$ increases,
$f$ increases. At one angle, (\cite{Sommer}), $f$ reaches its
maximum value
\begin{equation}\label{e:fmax}
f_{max}= \mu_s N.
\end{equation}
Beyond that angle, denoted as $\theta_0$, the pure rolling demands
value of $f$ more than $\mu_s N$. As this is not possible, the
small sphere starts to slip and roll simultaneously. In other
words, at that angle $\theta_0$, pure rolling stops. This feature
has been discussed well, in the context of motion of a cylinder on
a cylinder, in a beautiful paper of Flores et.al.(\cite{Flores}).
Along with it, as long as the small sphere pure rolls, there is no
loss of energy. The energy conservation equation (\ref{e:enr})
remains the same,
\begin{equation}
\frac{1}{2}mv^{2}+\frac{1}{2}I(w^{\prime})^{2}-mg(r+R)(1-cos\theta)=0
\end{equation}
This equation, in this case of pure rolling, is equivalent to
\begin{equation}\label{e:enrr}
\frac{v^{2}}{r+R}=\frac{10}{7}g (1-cos\theta).
\end{equation}
In the eq.(\ref{e:f}) at the angle, $\theta_0$, at which the small
sphere starts to slip, takes the form,
\begin{equation}
\mu_s N(\theta_0)=\frac{2}{7}mg sin\theta_0
\end{equation}
Consequently, the radial equation of motion (\ref{e:ra})
\begin{equation}
 N(\theta_0)=  mgcos\theta_0 - \frac{mv^2}{R+r}
\end{equation}
together with the energy conservation equation (\ref{e:enrr}),
leads to an implicit equation for $\theta_0$,
\begin{equation}\label{e:pha}
\mu_s = \frac{2sin\theta_0}{17cos\theta_0 -10}
\end{equation}
The plot against the static friction coefficient $\mu_s$, of the
angle, $\theta_0$, at which the small sphere starts slipping along
with rolling, is shown in the Fig.2. Interesting part of it is
that, the eq.(\ref{e:pha}) does not depend on any of the radii of
the two spheres involved.
\begin{figure}[ht]
\includegraphics{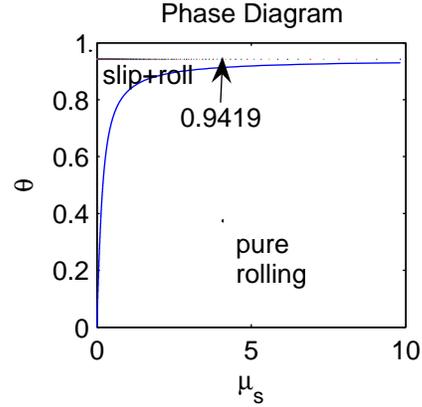}
\caption{\small{Plot of stopping angles for pure rolling against static friction coefficients}}\label{Figure2}
\end{figure}
The upper part of the phase diagram, in principle, can split into
two parts. One in which the mobile sphere remains on the static
sphere while doing mixed rolling. Another in which the mobile
sphere has left the surface of the static sphere. This second
line, obviously, will start from $\theta=cos^{-1}\frac{2}{3}$ and
will meet the first line (eq.\ref{e:pha}), at $\mu_{s}\rightarrow
\infty$
\subsection{Determination of angle of release, $\theta_{l}$}
Once the mobile sphere starts skidding, kinetic friction comes into play. The frictional force
is given by, (\cite{Sommer}),
\begin{equation}
f_{k} = \mu_{k} N
\end{equation}
where, $\mu_{k}$ is constant.
The cross radial equation of motion, (\ref{e:cra}), takes the form
\begin{equation}\label{e:mk}
ma_{\theta} = mg sin\theta - \mu_{k} N.
\end{equation}
This equation, eq.(\ref{e:mk}), combined with the radial equation, eq.(\ref{e:ra}), leads to
a first order differential equation for $\omega$,
\begin{equation}\label{e:int}
\frac{d\omega}{dt} -\mu_{k}\omega^{2} = \frac{g}{r+R}(sin\theta -\mu_{k}cos\theta).
\end{equation}
Integration, a.la Mele et.al.(\cite{Mele}) of the above equation,
results in an expression for $\omega$ for an angle greater than
the cross-over angle, $\theta_{0}$,\\
\noindent
$\omega^{2}(\theta)=\omega^{2}(\theta_{0})e^{2\mu_{k}(\theta-\theta_{0})}-\frac{2g}{(r+R)(1+4\mu_{k}^{2})}[cos\theta(1-2\mu_{k}^{2})\\
+3\mu_{k}sin\theta -(cos\theta_{0}(1-2\mu_{k}^{2})+3\mu_{k}sin\theta_{0})e^{2\mu_{k}(\theta-\theta_{0})}]$

\noindent
This expression when supplemented with the equations,
(\ref{e:normal}), (\ref{e:ra}), (\ref{e:enrr}), leads to an
illuminating relation,\\
\noindent
$3cos\theta_{l}+6\mu_{k}sin\theta_{l}= e^{2\mu_{k}(\theta_{l}-\theta_{0})}[6\mu_{k}sin\theta_{0}+\frac{4-68\mu_{k}^{2}}{7}\\
                                cos\theta_{0} +
                                \frac{10}{7}(1+4\mu_{k}^{2})].$
\noindent Once complemented with the equation, (\ref{e:pha}), this
is an expression for $\theta_{l}$ in terms  of $\mu_{s}$ and
$\mu_{k}$. One interesting simplification occurs for
$\mu_{k}=\mu_{s}$. Then as described just before this subsection,
it becomes crystal clear, see the Fig.\ref{Figure3}, that the
three phase lines separating pure rolling, mixed rolling, detached
phases, meet at one point. The point is
$\theta_{l}=cos^{-1}\frac{10}{17}=\theta_{0}$. This is reminiscent
of the triple point in the solid-liquid-gas phase transitions.
\begin{figure}[ht]
\includegraphics{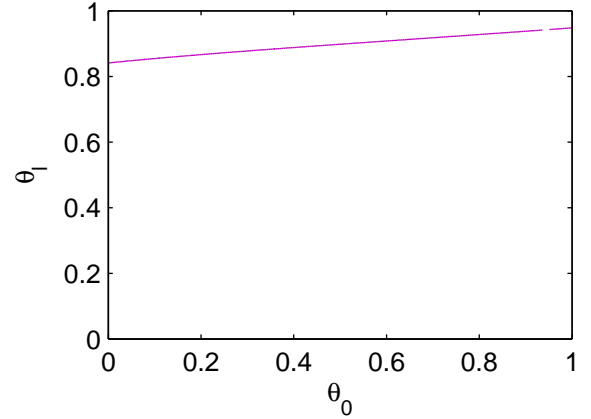}
\caption{\small{Plot of release angles against cross-over angles}}\label{Figure3}
\end{figure}
\subsection{Order parameter}
As long as the sphere pure rolls, $\frac{v}{\omega r}$ is one. As
it starts mixed rolling, $\frac{v}{\omega r}$ starts increasing
from one continuously. Hence, $2\pi(\frac{v}{\omega r}-1)$ acts
like an order parameter, $\eta$, of second order phase transition
\cite{{Alben}, {Jean}, {Fletcher}}. Here the phase transition is
intrinsically a non-equilibrium \cite{{Mario}, {Aubry}} one.
Physically, the introduced order parameter refers to the deficit
angle a point on the sphere is lagging by in terms of rotation per
second, in comparison to a hypothetical pure rolling sphere.
Analogues of $\Omega$ and $\Omega_{c}$ in Fig.3b. of
ref.(\cite{Jean}) are the static frictions given by the equations,
(\ref{e:f}) and (\ref{e:fmax}). We observe,
\begin{equation}
\eta
=\pi\mu_{k}(5+\frac{2-\frac{5\mu_{s}}{\mu_{k}}+cos\theta_{0}(\frac{17}{2}\frac{\mu_s}{\mu_k}-\frac{11}{2})}
{1-cos\theta_{0}}) \frac{sin\theta_{0}}{cos\theta_{0}}
(1-\frac{f_{max}}{f})
\end{equation}
Hence, we get a critical exponent $1$.
\end{section}
\begin{section}{Acknowledgement}
The authors would like to thank Anupam and Saiteja for help with
plotting.
\end{section}
\begin{section}{Suggested Problems}
The three dimensional, ($\theta_{l},\mu_{s},\mu_{k}$), phase
diagram can be easily drawn using Mathematica7. The phase diagram
for the case when the small sphere starts with an initial kinetic
energy, has been done for $\frac{v_{0}}{\omega_{0}r}=1$ in ref.
\cite{Puneet}. Moreover, it will be interesting to have a full
qualitative analysis of motion of a sphere on a static sphere
following Flores et.al.(\cite{Flores}). Exploring the consequences
of replacing the static sphere by, say, a static cycloidal hill
(\cite{Spiegel}), along the line of this article, is as much
desirable.
\end{section}

\end{document}